\pdfoutput=1

\documentclass[11pt]{article}

\usepackage[final]{acl}
\usepackage{makecell}
\usepackage{multirow}
\usepackage{times}
\usepackage{latexsym}
\usepackage{amsmath}
\usepackage{amssymb}
\usepackage{arydshln}
\allowdisplaybreaks
\usepackage[T1]{fontenc}

\usepackage[utf8]{inputenc}

\usepackage{microtype}

\usepackage{inconsolata}
\usepackage{tabularx}
\usepackage{graphicx}
\PassOptionsToPackage{table}{xcolor}
\usepackage{colortbl}
\usepackage{color}
\definecolor{shadecolor}{rgb}{0.9,0.9,0.9}
\usepackage{tabularx}
\usepackage{booktabs}
\usepackage{verbatim}
\usepackage{utfsym}
\usepackage{xurl}

\title{Adapting General-Purpose Embedding Models to Private Datasets\\ Using Keyword-based Retrieval}

\author{Yubai Wei, Jiale Han \and Yi Yang \\
  Hong Kong University of Science and Technology \\ 
  \texttt{yubaiwei@ust.hk}, \texttt{jialehan@ust.hk}, \texttt{imyiyang@ust.hk}
 }

\begin{document}
\maketitle
\begin{abstract}
Text embedding models play a cornerstone role in AI applications, such as retrieval-augmented generation (RAG). While general-purpose text embedding models demonstrate strong performance on generic retrieval benchmarks, their effectiveness diminishes when applied to private datasets (e.g., company-specific proprietary data), which often contain specialized terminology and lingo. In this work, we introduce BMEmbed, a novel method for adapting general-purpose text embedding models to private datasets. By leveraging the well-established keyword-based retrieval technique (BM25), we construct supervisory signals from the ranking of keyword-based retrieval results to facilitate model adaptation. We evaluate BMEmbed across a range of domains, datasets, and models, showing consistent improvements in retrieval performance. Moreover, we provide empirical insights into how BM25-based signals contribute to improving embeddings by fostering alignment and uniformity, highlighting the value of this approach in adapting models to domain-specific data. We release the source code\footnote{The code is available at: \url{https://github.com/BaileyWei/BMEmbed}.} for the research community.

\end{abstract}
\section{Introduction}

Text embeddings serve as a cornerstone for various AI applications, particularly in information retrieval and retrieval-augmented generation (RAG) systems \cite{DBLP:journals/tmlr/IzacardCHRBJG22, DBLP:journals/corr/abs-2312-10997}. With the widespread adoption of AI, companies like OpenAI and Cohere now provide general-purpose text embedding APIs, enabling organizations to quickly integrate AI into their RAG systems. However, while these general-purpose embedding models show impressive performance on generic benchmarks, they often face significant challenges when applied to private datasets, such as domain-specific or company-specific proprietary data, which often contain specialized terminology and jargon \cite{DBLP:conf/emnlp/AndersonJHCF24, DBLP:journals/corr/abs-2409-18511}.

For instance, consider a pharmaceutical company that seeks to build a RAG system over its vast internal dataset. The company’s employees may query the system for information about an internal product code (e.g., Product Code: PHX-121). However, general-purpose models, not trained on this proprietary dataset, may fail to properly interpret or retrieve relevant documents containing such specific terms, leading to suboptimal answers. 

\begin{figure}[t!]
	\centering
	\includegraphics[width=1\linewidth]{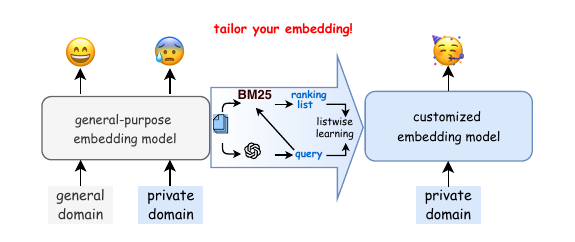}
	\caption{An illustration of tailoring an embedding model to a private domain.}
	\label{ointro} 
\end{figure}

Current practices in RAG systems often attempt to address this issue by combining traditional keyword-based retrieval with embedding-based retrieval. One popular hybrid approach is reciprocal rank fusion (RRF), which reranks results based on a mathematical formula without fine-tuning the underlying embedding model \citep{DBLP:conf/sigir/CormackCB09}. While simple and effective, RRF remains heuristic, with its effectiveness potentially limited by the lack of fine-tuning to the private dataset. 
This leads us to the following question: \textit{Can we fine-tune general-purpose embedding models to better align with private datasets?}

One of the key challenges in adapting embedding models to domain-specific datasets is the lack of available tuning signals. While general-purpose embedding models are often trained on large, curated QA datasets using contrastive learning \cite{DBLP:conf/acl/TanSWYS22, DBLP:conf/acl/ZhouZZW22, moreira2024nv}, private datasets, which often consist of free-text data without annotations, pose a particular challenge. This leads to an important sub-question: \textit{How can we generate supervisory signals for adapting general-purpose embedding models to private, unlabeled datasets?}

In this work, we introduce BMEmbed, an automated framework designed to adapt general-purpose text embedding models to private datasets. Our method leverages BM25 \citep{DBLP:journals/ftir/RobertsonZ09}, a well-established keyword-based retrieval function based on TF-IDF, to generate supervisory signals from the ranking of keyword-based retrieval results. The BMEmbed framework consists of three main components: (1) domain query generation, where a large language model generates synthetic queries based on domain-specific events extracted from the private corpus; (2) relevant sampling, which uses BM25 to retrieve lexically related paragraphs and samples from different intervals of the ranking list to ensure informative signals; and (3) listwise fine-tuning, where the embedding model is optimized using a listwise loss function on the curated ranking lists, fully leveraging the ranking supervision. Unlike traditional in-batch negative contrastive learning \citep{DBLP:journals/corr/abs-1807-03748, DBLP:conf/icml/ChenK0H20}, our approach uses ranked BM25 results to  guide the fine-tuning process.

We evaluate BMEmbed across multiple domains and datasets, using two general-purpose embedding models with varying scales. Compared to base embedding models, BMEmbed consistently achieves substantial improvements in retrieval accuracy. Our experiments further show that BMEmbed outperforms or achieves competitive performance compared to two commonly used techniques in current RAG systems: (1) fine-tuning with in-batch negative contrastive learning, and (2) the RRF hybrid approach. 
To better understand the inner workings of BMEmbed, we investigate the alignment and uniformity properties of the adapted embeddings \citep{DBLP:conf/icml/0001I20}. We find that BMEmbed successfully improves embedding uniformity while maintaining good alignment, leading to improved retrieval performance. 

In summary, this paper introduces a simple yet effective method for adapting general-purpose text embedding models to private datasets. Given the increasing adoption of RAG systems across industries, we believe our method provides a practical solution to enhance domain specificity, leading to more accurate and contextually relevant retrieval results in real-world applications.

\section{Background}

\subsection{Text Embedding Models}

Text embedding refers to the numerical representation of a piece of text that captures its semantic meaning, transforming texts of varying lengths into fixed-size vectors. Previously, fine-tuning models like BERT \cite{DBLP:conf/naacl/DevlinCLT19} and T5 \cite{DBLP:journals/jmlr/RaffelSRLNMZLL20} to adapt to embedding downstream tasks was the dominant approach \cite{DBLP:conf/emnlp/ReimersG19,DBLP:conf/acl/NiACMHCY22}. However, with the development of LLMs, the landscape is shifting. The focus has now moved toward building LLM-based, general-purpose embedding models, including Qwen \cite{DBLP:journals/corr/abs-2308-03281}, LLM2Vec \cite{DBLP:journals/corr/abs-2404-05961}, NV-Embed \cite{DBLP:journals/corr/abs-2405-17428}, etc. These LLM-based embedding models have demonstrated their superiority on massive text datasets, e.g., MTEB \cite{DBLP:conf/eacl/MuennighoffTMR23}.

Current embedding models \citep{DBLP:journals/tmlr/IzacardCHRBJG22, DBLP:journals/corr/abs-2212-03533, DBLP:journals/corr/abs-2308-03281, DBLP:conf/acl/ChenXZLLL24,tang2024pooling} are primarily trained using contrastive learning, with the widely adopted InfoNCE loss\cite{DBLP:journals/corr/abs-1807-03748} as the objective, which aims to distinguish semantically relevant text pairs from irrelevant ones. 

While effective, the performance of contrastive learning heavily depends on the selection of high-quality positive and negative samples \citep{DBLP:conf/acl/TanSWYS22, DBLP:conf/acl/ZhouZZW22, moreira2024nv}. When adapting the embedding model to a specific domain, constructing relevant and irrelevant samples from a private corpus can be a challenging task. In this work, we propose leveraging BM25 to construct lexically relevant samples, addressing the challenge of sample selection in an unsupervised manner.

\subsection{Keyword-based Retrieval: BM25}

\begin{figure*}[t]
\centering
\includegraphics[width=0.8\textwidth]{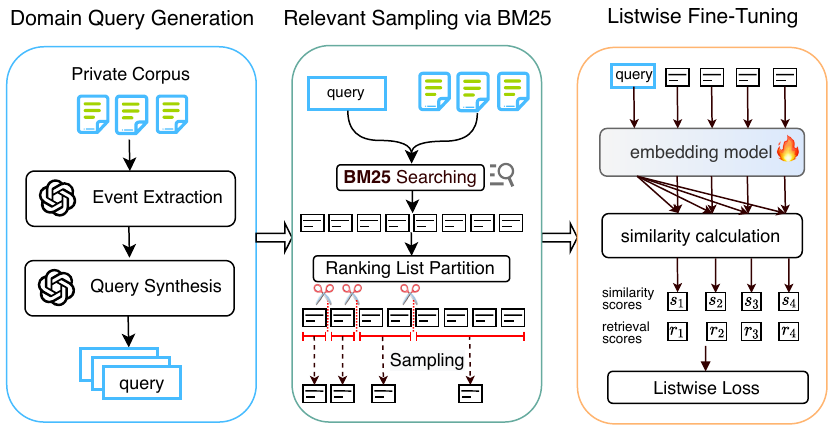}
  \caption {An overview of the BMEmbed framework.} 
   \label{tab:pipeline}
\end{figure*}

BM25 \cite{DBLP:journals/ftir/RobertsonZ09}is a well-established retrieval method based on TF-IDF, which ranks documents by considering the uniqueness and significance of terms relevant to a given query. The BM25 score for document $d$ with respect to query $q$ is defined as:
\begin{equation*}
 \small
  \label{eq:contrastive loss}
  \begin{split}
\text{BM25}(d, q) = \sum_{t \in q} \text{IDF}(t) \cdot \frac{f(t, d) \cdot (k_1 + 1)}{f(t, d) + k_1 \cdot \big(1 - b + b \cdot |\hat{d}|\big)}
\end{split}
\end{equation*}
where \(f(t, d)\) is the term frequency of term \(t\) in document \(d\), \(|\hat{d}|\) is the normalization of document length, \(\sum_{t \in q} \text{IDF}(t)  \) is the inverse document frequency of term \(t\) in the corpus, \(k_1\) and \(b\) are hyper parameters that control the impact of term frequency and document length, respectively. Previous works have demonstrated the effectiveness of using BM25 as a weak supervision signal for training small models \citep{DBLP:conf/sigir/DehghaniZSKC17, DBLP:conf/sigir/HaddadG19,DBLP:conf/emnlp/KarpukhinOMLWEC20}.

Despite significant progress in dense retrieval \cite{DBLP:conf/emnlp/KarpukhinOMLWEC20, DBLP:conf/acl/XinX0SJ022}, BM25 remains a robust retrieval algorithm. Its rule-based, keyword matching approach enables strong generalization, maintaining competitive performance in scenarios where keyword matching is more crucial than semantic matching. As a result, hybrid approaches, such as Reciprocal Rank Fusion (RRF) \citep{DBLP:conf/sigir/CormackCB09}, have been used to combine and rerank results from both dense retrieval models (embedding-based) and sparse retrieval models (BM25-based). However, RRF relies on heuristics to rank these hybrid results. In contrast, this paper aims to fine-tune general-purpose embedding models to a specific dataset, enabling true adaptation rather than simply combining results from different retrieval methods.

\section{BMEmbed: Domain Adaptation for General-Purpose Embeddings}
In this section, we present BMEmbed, an automated framework designed to tailor general-purpose embedding models to private datasets consisting of unannotated text.  
The method contains three steps, and the overall process is illustrated in Figure~\ref{tab:pipeline}. 

\subsection{Domain Query Generation}\label{Domain_Query_Generation}
The first step is to prompt an LLM (e.g., GPT-4) to generate synthetic queries focused on domain-specific events in the private corpus, rather than on general concepts.

\paragraph{Event Extraction} We require the LLM to extract all the events and their associated arguments from the private corpus. In addition, the original context from which the events are extracted is also generated, serving as the evidence for the queries used in the baseline method in subsequent experiments.

\paragraph{Query Synthesis} Then, we feed both the corpus and the extracted events into the LLM, prompting it to automatically generate queries \(Q\) for each event. The detailed prompts are provided in Appendix~\ref{sec:appendix_prompt}.

\subsection{Relevant Sampling via BM25}

The second step is to construct ranked retrieval results using keyword retrieval method BM25. 

\paragraph{BM25 Searching} We divide the private corpus into multiple chunks and calculate the BM25 score between query $q\in Q$ and each chunk. The top-\( k \) scoring chunks, denoted as 
\( C = \{c_1, c_2, \dots, c_k\} \), are selected, where each chunk \( c_i \) is associated with its respective BM25 score \( r_i \).

\paragraph{Ranking List Partition} We further partition \( C \) into \( m \) intervals, denoted as \( \{ \mathcal{P}_1, \mathcal{P}_2, \dots, \mathcal{P}_m \}\). This approach allows positives and negatives to be sampled from different intervals, which amplifies the scope of sampling space across diverse relevance tiers, effectively mitigating noise in BM25 pseudo labels. The partitioning can follow either a uniform or a fine-to-coarse strategy. Uniform intervals divide the range of BM25 scores into equally sized segments, ensuring a consistent distribution of samples across all intervals. In contrast, fine-to-coarse partitioning strategy intervals prioritize finer segmentation of higher-relevance scores, leading to more granular sampling for positively ranked examples. 
For instance, given \(m=4\), the top-20 ranking list can be divided into intervals \([0,2),[2,6), [6, 12),[12, 20)\) using the fine-to-coarse strategy, whereas the uniform strategy divides it into \([0,5),[5,10), [10, 15), [15, 20)\).

\paragraph{Ranking-Based Sampling} For each interval \( \mathcal{P}_j \), we randomly select one sample \( p_j \) along with its retrieval score \( r_j \), forming a ranking list \( [q, p_1, p_2, \dots, p_m, r_1, r_2, \dots, r_m] \). 

\subsection{Listwise Fine-Tuning}
Since BM25 retrieval results produce a ranked list, we hypothesize that this ranking contains valuable information that can be better utilized through a listwise training objective, rather than the commonly used in-batch negative contrastive learning objective, where ranking information is typically ignored. To this end, we employ a listwise training objective to fully leverage the ranking information obtained from BM25 retrieval. 

Given \( [q, p_1, p_2, \dots, p_m, r_1, r_2, \dots, r_m] \) and a base embedding model $e(\cdot)$, we first obtain the embeddings of $q$ and \( p_j \) for \( j \in [1, \dots, m] \), denoted as \( e(q) \) and \( e(p_j) \), respectively. Then, we calculate the cosine similarity \( s_j=\text{sim}(e(q), e(p_j)) \). Following the work of ListNet \citep{DBLP:conf/icml/CaoQLTL07}, the listwise loss is calculated as follows:
\begin{equation*}
  \label{eq:cross entropy}
  \mathcal{L}(s, r) = -\sum_{q \in Q} \sum_{j=1}^{m} p^{r}_j \log(p^{s}_j)
\end{equation*}
where $r$=$\{r_1, r_2, \dots, r_m\}$, $s$=$\{s_1, s_2, \dots, s_m\}$, \( p^{r} \) and \( p^{s} \) are the distributions normalized by softmax over the $r$ and $s$, respectively. We introduce a temperature scaling factor \( \alpha \) on the target score list \( r \), with:
\begin{equation*}
p^r_j = \frac{\exp\left(\frac{r_{j}}{\alpha}\right)}{\sum_{i=1}^{m} \exp\left(\frac{r_{i}}{\alpha}\right)}
\end{equation*}
Here, \( \alpha \) controls the sharpness of the target distribution, with smaller values leading to a more concentrated distribution, and larger values resulting in a smoother distribution.

\section{How does BMEmbed Perform?}

\subsection{Experimental Setup}
\paragraph{Base Embedding Models} We use the following two general-purpose embedding models: gte-Qwen2-1.5B-instruct\footnote{\scriptsize \url{https://huggingface.co/Alibaba-NLP/gte-Qwen2-1.5B-instruct}}, a small yet strong model, and e5-mistral-7B-instruct\footnote{\scriptsize \url{https://huggingface.co/intfloat/e5-mistral-7b-instruct}}, a larger model based on Mistral-7B. Both two models perform competitively on the MTEB leaderboard \cite{DBLP:conf/eacl/MuennighoffTMR23}.

\begin{table}[t]
	\centering
	\renewcommand\tabcolsep{2.8pt}
	\scalebox{0.88}
	{
		\begin{tabular}{lccc}  
			\toprule
			 
			Dataset &Multihop&Finance&LegalBench\\\midrule
            evaluation queries & 2,255& 498& 1,676\\
            corpus tokens& 1,453k& 840k& 7,109k\\
            synthesized queries& 5,972& 1,009& 685\\
            chunk size & 256& 1,024& 1,024\\
            $k$ & 1,000& 1,000& 4,000\\
            $m$ & 9& 6& 6\\
			\bottomrule
		\end{tabular}
	}
	\caption{
		Statistics and implementation details of the datasets.}
	\label{Statistics}
\end{table}

\paragraph{Baselines} We compare models fine-tuned by BMEmbed with the following methods: 1) \textbf{BM25}, with parameters $k_1$=1.2 and $b$=0.75; 2) \textbf{Base}, the base embedding model. 3) \textbf{CL}, the embedding model fine-tuned using contrastive objective InfoNCE loss \cite{DBLP:journals/corr/abs-1807-03748}, where LLM-generated evidence is used as positives (as detailed in Section~\ref{Domain_Query_Generation}), along with in-batch negatives. 4) \textbf{RRF}, Reciprocal Rank Fusion \cite{DBLP:conf/sigir/CormackCB09}, which is a hybrid search method combining rankings from multiple sources into a unified ranking:
\begin{equation*}
\small
RRF(d)=\sum \limits_{a \in A}\frac{1}{u+a(d)}
\end{equation*} where $d$ is a document, $A$ is the set of rankers (retrievers), $a(d)$ is the rank of document $d$ in ranker $a$, and $u$ is a constant set to 40. Here we combine BM25 rankings with the base embedding model. 5) \textbf{RRF+BMEmbed}, the combination of the BM25 and the BMEmbed-finetuned model.

\begin{table*}[t]
  \centering
  \resizebox{1\linewidth}{!}{
    \begin{tabular}{l|cccc|cccc|cccc}
      \toprule
      \multirow{1}*{Method} & \multicolumn{4}{c|}{Multihop-RAG} & \multicolumn{4}{c|}{Finance-RAG} & \multicolumn{4}{c}{LegalBench-RAG} \\
      & Hit@1 & Hit@4 & Hit@10& MAP@10 &Hit@1 & Hit@4 & Hit@10& MAP@10 &Hit@1 & Hit@4 & Hit@10& MAP@10 \\
          \midrule
      BM25& 41.06& 65.01&79.02& 25.93&28.51 & 46.18&57.43&  37.46&0.12  &  7.58& 14.62& 1.62\\
        \midrule
\rowcolor{shadecolor}
 \multicolumn{13}{c}{Qwen2-1.5B}\\
 Base& 33.97 & 59.69& 76.50& 22.22&23.69 &41.37& 53.82& 32.84& 8.00&16.65&23.09&6.34\\
CL & 31.53& 55.96&74.72& 21.48& 25.50 & 43.57&58.43& 35.20& 6.44 &  17.90& 25.48& 5.45\\
 \textbf{BMEmbed}& 40.58 &68.34&83.06& 26.54&26.31 & 45.38&57.03&36.21& 8.95&\textbf{20.64}&\textbf{28.52}&\textbf{7.47}\\       
 RRF& 38.76& 66.30&82.04& 25.80& \textbf{31.73}& 49.80&63.45& 40.97& 8.47 & 18.32&24.76&  6.45\\
  RRF+\textbf{BMEmbed}& \textbf{43.28}& \textbf{71.09}& \textbf{84.35}&\textbf{28.30}& \textbf{31.73}&\textbf{51.61} &\textbf{64.46}& \textbf{41.62}&\textbf{9.43} & 19.69&28.46&7.19\\
      \midrule
\rowcolor{shadecolor}
       \multicolumn{13}{c}{e5-mistral-7B}\\
 Base& 29.49& 54.99&75.39& 20.33& 19.28& 36.55&48.80& 28.10&7.76 & 17.42&23.75&6.48\\
 CL & 21.11& 48.34&69.40& 16.67& 24.30& 46.79&57.43& 35.08& 7.88 &  16.65&21.06&  5.37\\
 \textbf{BMEmbed} & 45.28 & \textbf{71.49}&85.63& 27.60&28.11 & 48.39&62.25& 38.40& \textbf{9.96}& 19.03&\textbf{27.27}&7.08\\   
 RRF& 42.26& 67.58&82.13& 27.04& 30.72& 47.39&61.85& 39.55& 9.79&  \textbf{19.09}&24.34& \textbf{7.23}\\
 RRF+\textbf{BMEmbed}& \textbf{45.72} & 71.44&\textbf{85.72}&  \textbf{28.36}& \textbf{32.33}& \textbf{52.21}&\textbf{64.06}& \textbf{41.92}& \textbf{9.96}&19.03&\textbf{27.27}&7.08\\

 \bottomrule
    \end{tabular}
}
  \caption{Retrieval performance of different methods across three datasets. \textbf{Best} results are highlighted for each embedding model on each dataset.}
  \label{main results}
\end{table*}

\paragraph{“Private” Datasets} In our experiments, we choose three publicly available retrieval datasets as evaluation benchmarks. However, these datasets are released after the base embedding models, meaning the models are unlikely to have been trained on them. Therefore, while the datasets are publicly available, they effectively simulate “private” datasets in our experiments, also ensuring fair comparison and reproducibility.

Specifically, the three datasets are: Multihop-RAG \cite{DBLP:journals/corr/abs-2401-15391}, a multi-hop question answering (QA) dataset from the financial news domain; 
Finance-RAG\footnote{\scriptsize \url{https://www.kaggle.com/competitions/icaif-24-finance-rag-challenge}}, a long-context QA dataset based on financial reports, released as part of the ACM-ICAIF'24 FinanceRAG competition; 
and LegalBench-RAG \cite{DBLP:journals/corr/abs-2408-10343}, a challenging long-context legal domain QA dataset. 
Each dataset contains questions, their corresponding relevant evidence, and the original corpus. We use the evidence as the label to evaluate the retrieval performance. Detailed statistics are provided in Table~\ref{Statistics}.

\paragraph{Implementation and Training Details}  
For domain query generation, we use GPT-4o for accurate event extraction and GPT-4o-mini for query synthesis to minimize costs. We generate 5,972, 1,009, and 685 queries for Multihop-RAG, Finance-Bench, and Legal-Bench, respectively, based on corpus size. A real case, including the input corpus, intermediate events, and the final generated query, is showcased in Appendix~\ref{case_query_generation}.  During relevant sampling, we set the chunk size of 256 for Multihop-RAG and 1,024 for the other two datasets with long context. The fine-to-coarse partitioning strategy is used by default. We adopt \(m\)=9 for Multihop-RAG and \(m\)=6 for the others,  with $k$=1,000 for MultiHop-RAG and Finance-RAG, and $k$=4,000 for LegalBench-RAG. The impact of different $m$ and partitioning strategies is further discussed in Section~\ref{effect_Partitions}. The results under different $k$ are shown in Appendix~\ref{app:ablation}. For finetuning, we use a fixed batch size of $16$ for CL, while the batch size is equivalent to \(m\) for BMEmbed. The temperature $\alpha$ is set to a moderate value between 1.0 to 5.0, with further adjustments on different datasets and models, which we provide a detailed discussion in Section~\ref{effect_Temperatures}. We finetune the model using LoRA \citep{DBLP:conf/iclr/HuSWALWWC22} with a rank of 16 for 1,000 steps. Training Qwen on 4$\times$3090 GPUs takes about 1.5 hours, while training e5-mistral on 8$\times$H800 GPUs takes approximately one hour.

\subsection{Results and Discussion}

Table \ref{main results} presents the experimental results of BMEmbed and all baselines across two embedding models and three datasets. It can be observed: 

1) The vanilla embedding models perform suboptimally in specific domains. In most cases, base models underperform BM25 on Multihop-RAG and Finance-RAG, even with large model sizes. This finding highlights the necessity of further adaptation when applying general-purpose embedding models to specific domains. Furthermore, the BMEmbed consistently outperform BM25 across models and datasets, despite being trained with supervisory signals derived from BM25. This demonstrates that BMEmbed is not merely mimicking BM25. Instead, we treat BM25 as a weak lexical teacher and design both our sampling strategy and training objective to guide the model toward learning relevance information beyond BM25's direct outputs.

2) Contrastive learning does not consistently lead to performance improvements for embedding model adaptation. Surprisingly, we find that applying CL to base models do not always improve performance. 
We hypothesize that noise in the positive evidence generated by the LLM might interfere with model optimization. This indicates that contrastive learning is sensitive to the quality of positive and negative samples, and such an approach does not always result in promising improvements for embedding adaptation.

\begin{figure}[t]
\centering
\includegraphics[width=0.8\linewidth]{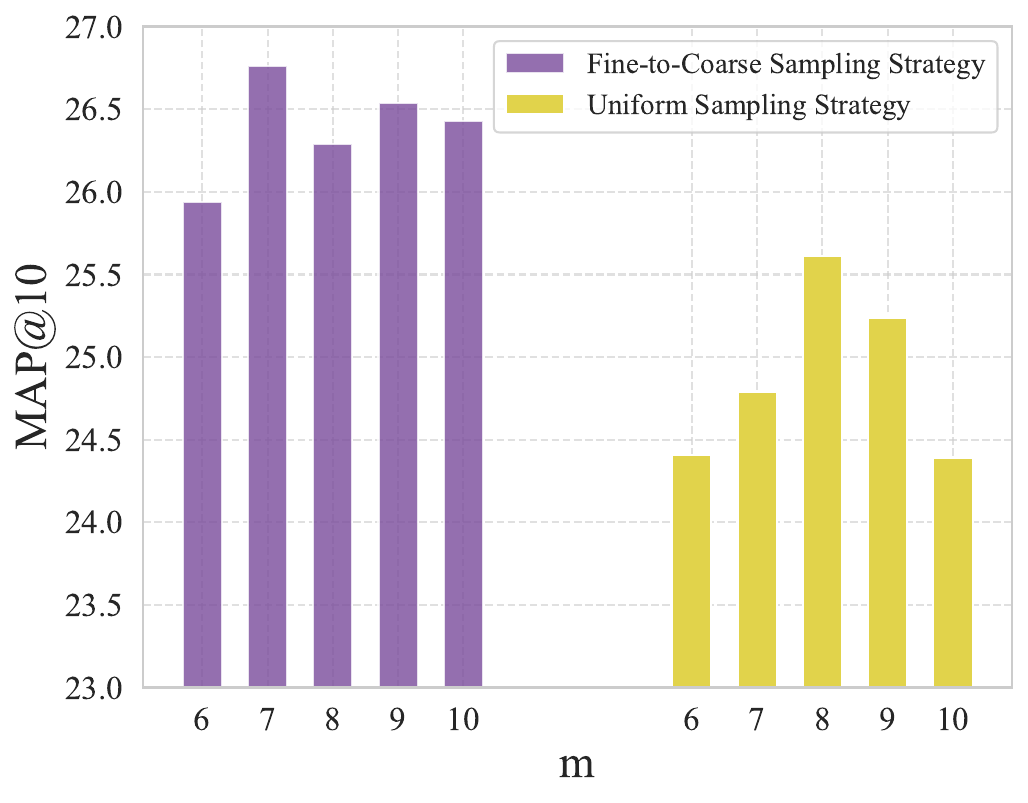}
  \caption {Retrieval performance of MAP@10 for different $m$ and sampling strategies.} 
   \label{tab:partition retrieval}
\end{figure}

\begin{figure}[t]
\centering
\includegraphics[width=0.8\linewidth]{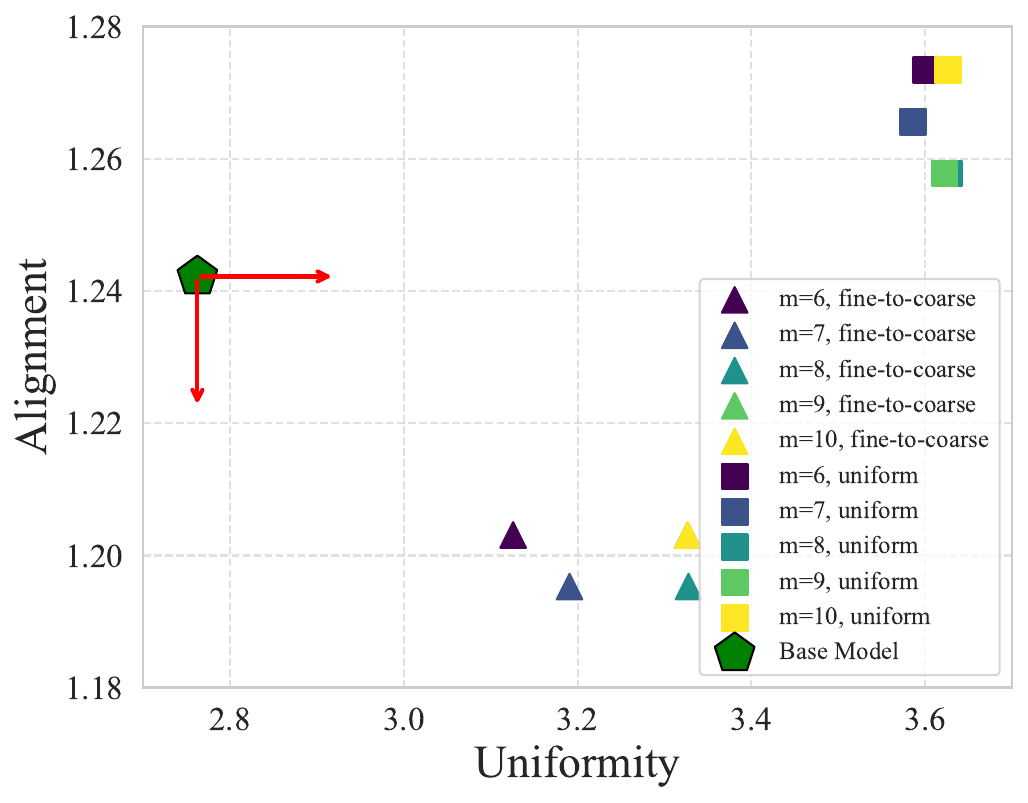}
  \caption {Alignment and uniformity for different $m$ and sampling strategies.} 
   \label{tab:sampling strategy}
\end{figure}

\begin{figure}[t]
\centering
\includegraphics[width=0.76\linewidth]{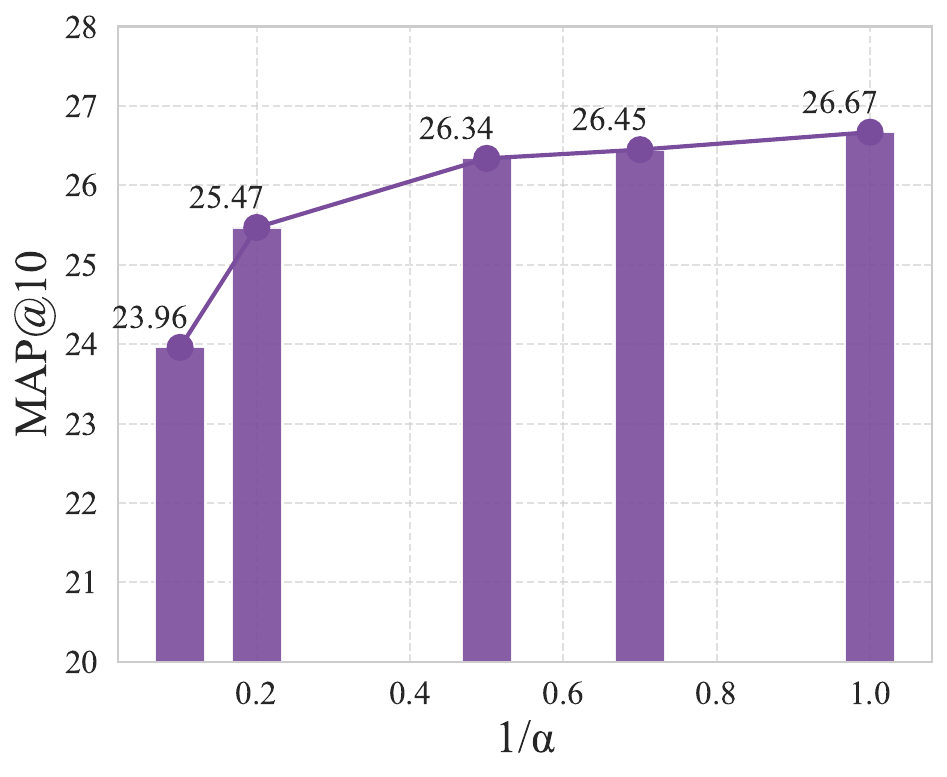}
  \caption {
  Retrieval performance of MAP@10 for different $\alpha$.
  } 
   \label{tab:retrieval and temperature}
\end{figure}

\begin{figure}[t]
\centering
\includegraphics[width=0.8\linewidth]{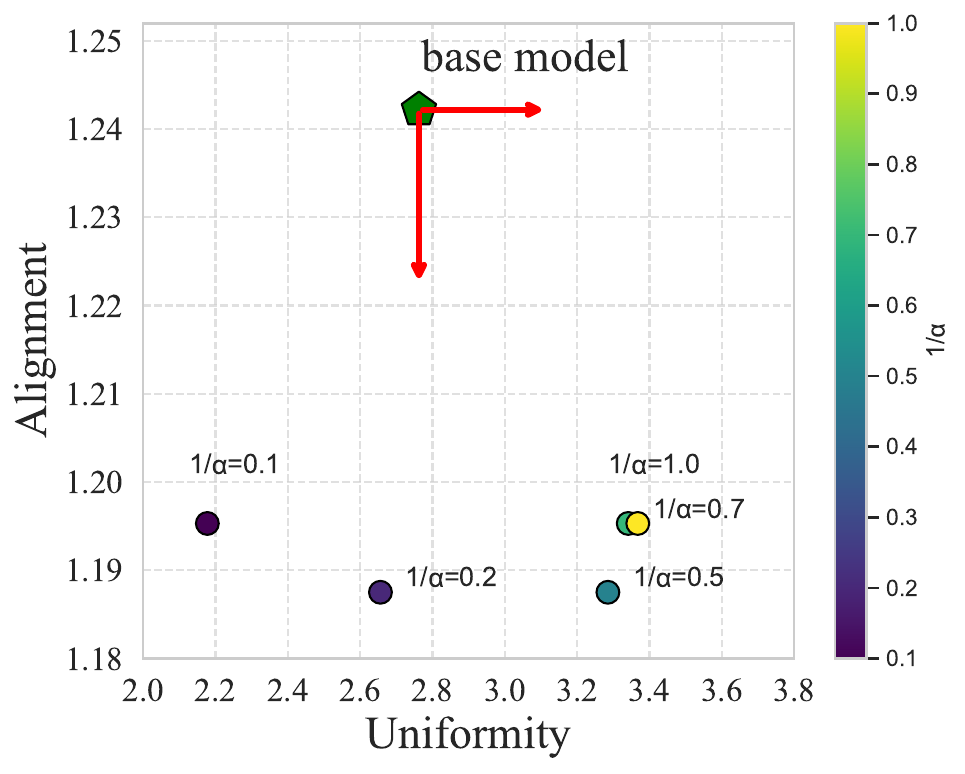}
  \caption {
  Alignment and uniformity for different $\alpha$.
  } 
   \label{tab:alpha}
\end{figure}

3) Our BMEmbed consistently delivers improvements, benefiting from the supervision signals provided by BM25. Our framework boosts the base models across all embedding models and datasets, especially on the metrics Hit@4. Compared to RRF which combines BM25 ranking information with dense retrieval from embedding models, BMEmbed achieves a remarkable improvement, which illustrates that our framework deeply deciphers the ranking confidence signals from BM25, achieving a better embedding model adaptation.

4) Furthermore, BMEmbed can be combined with other hybrid retrieval methods to achieve further enhancement. This is demonstrated in experiments comparing RRF+BMEmbed with RRF alone. In most cases, RRF+BMEmbed shows clear performance gains, except in the case of LegalBench-RAG, where the BM25 baseline performs poorly and BMEmbed+RRF does not achieve further performance gains.

\label{main experiment}

\subsection{Generality under Alternative Settings}
To further explore the generality of the BMEmbed framework, we conduct additional experiments under three different settings: (1) applying BMEmbed to a smaller embedding model, (2) replacing the loss function in listwise fine-tuning, and (3) evaluating the adapted embedding model on other embedding task. Full experimental setups and results are provided in Appendix~\ref{appendix:generalization}. 

In \textbf{Setting 1}, we choose all-MiniLM-L6-v2\footnote{\scriptsize \url{https://huggingface.co/sentence-transformers/all-MiniLM-L6-v2}} from the Sentence Transformers family as a smaller embedding model. We observe that even small model can achieve performance comparable to larger general-purpose model after BMEmbed adaptation, while requiring significantly fewer computational resources and training time. This highlights the practicality and efficiency of our framework in resource-constrained scenarios. In \textbf{Setting 2}, we replace the cross-entropy loss with a maximum likelihood loss, ListMLE \cite{xia2008listwise} in listwise fine-tuning. The adapted model still shows consistent improvements, demonstrating that BMEmbed is robust across different listwise training objectives. In \textbf{Setting 3}, we evaluate the adapted embedding Qwen2-1.5B on FinSTS\cite{liu2024beyond}, a semantic textual similarity task. Despite being trained solely on listwise signals derived from BM25 rankings and without any direct supervision on the STS task, the adapted models achieve noticeable improvements. This suggests that our approach effectively captures domain-specific semantic nuances, further highlighting its broader utility.

\begin{table*}[t!]
  \centering
  \resizebox{0.9\linewidth}{!}{
    \begin{tabular}{l|cc|cc|cc}
      \toprule
      \multirow{2}*{Method} & \multicolumn{2}{c|}{Multihop-RAG} & \multicolumn{2}{c|}{Finance-RAG} & \multicolumn{2}{c}{LegalBench-RAG} \\
      & Alignment$\downarrow$ & {Uniformity$\uparrow$} & Alignment$\downarrow$ & {Uniformity$\uparrow$}  & Alignment$\downarrow$ & {Uniformity$\uparrow$} \\
          \midrule
\rowcolor{shadecolor}
 \multicolumn{7}{c}{Qwen2-1.5B}\\
 Base & 1.2422& 2.7665& 1.1562& 1.6567 & \textbf{1.3203}&1.1599\\
 CL & 1.3516 & 2.8022 & 1.2188 & \textbf{2.9437} & 2.0000& \textbf{2.2382}\\
 BMEmbed & \textbf{1.2031} & \textbf{3.3266} & \textbf{1.1484} & 2.6631 & 1.6691 &2.1426\\   
      \midrule
\rowcolor{shadecolor}
\rowcolor{shadecolor}
       \multicolumn{7}{c}{e5-mistral-7B}\\
      Base& 1.1875& 1.7430& 1.1797& 1.0353& \textbf{1.2891}&0.7317\\
      CL& 1.5156 & 2.7649& 1.3281& 3.0445& 2.7969&\textbf{1.7913}\\
     BMEmbed & \textbf{1.1797}& \textbf{3.7768} & \textbf{1.0859}& \textbf{3.2144}& 1.6797&1.6182\\      
 \bottomrule
    \end{tabular}
}
\caption{Alignment and Uniformity of Embedding Models. Lower alignment ($\downarrow$) and higher uniformity ($\uparrow$) are preferred. \textbf{Best} results are highlighted for each embedding model on each dataset.}
  \label{main results 2}
\end{table*}

\section{Why BMEmbed Enhances Embedding Adaptation? An Investigation from Uniformity and Alignment} \label{uniformity and alignment}
In this section, we further investigate why BMEmbed leads to improvements. We conduct ablation experiments to study how our samplers and temperature interact with retrieval performance. Moreover, we introduce the \textbf{Alignment} and \textbf{Uniformity} properties, which reflect the quality of the embedding, to gain a deeper theoretical understanding. The reported experiments are based on the Multihop-RAG dataset and the Qwen2-1.5B model by default. The complete ablation study setup and results are presented in Appendix \ref{app:ablation}. As observed in the ablation study, our experiments empirically reveal \textbf{a strong agreement between embedding properties and retrieval performance, suggesting that the enhancement from BMEmbed results from the optimized embedding properties.} Here, we discuss our key observations and conclusions.

\subsection{Alignment and Uniformity}

A good embedding should bring similar data points closer together while preserving as much useful information as possible \cite{DBLP:conf/nips/BachmanHB19,DBLP:conf/iclr/HjelmFLGBTB19} to distinguish different data points, leading to \textit{lower alignment and higher uniformity}. Here, we adopt alignment and uniformity for evaluating an embedding following the work of \citet{DBLP:conf/icml/0001I20}, with further details and discussion in Appendix \ref{app:alignment_uniformity}. 

\subsection{Ablation Study of Different Partitions}\label{effect_Partitions}

To explore the effect of different partitions during relevant sampling via BM25 in BMEmbed, we investigate the impact of various partition factors, including the number of partitions and the partitioning strategies. Specifically, we conduct experiments with $m$ ranging from 6 to 10, using both uniform and fine-to-coarse sampling strategies, with the temperature $\alpha$ set to 1 and $k$ set to 1,000. 

Figure~\ref{tab:partition retrieval} shows the relationship between retrieval metrics MAP@10 and fine-tuning with different m and sampling strategies, while Figure~\ref{tab:sampling strategy} presents a comparison of uniformity and alignment of the fine-tuning models shown in previous figure. We observe that \textbf{the fine-to-coarse strategy achieves better retrieval performance and superior alignment compared to the uniform strategy}. In contrast, the uniform strategy is suboptimal in retrieval performance due to its overly uniform embedding distribution, which leads to a loss of alignment. In addition, as $m$ increases from 6 to 7 under the fine-to-coarse sampling strategy, we observe a measurable improvement in MAP@10 performance, suggesting that moderately expanding the sampling scope captures more relevant items. However, further increasing $m$ causes performance fluctuations and a gradual decline in overall effectiveness. These findings highlight the importance of carefully calibrating $m$ to optimize retrieval performance.

\begin{table*}[ht]
  \centering
  \resizebox{0.9\textwidth}{!}{
  \begin{tabularx}{\textwidth}{X|X|X}  
    \toprule
    Original Query& Masked Query& Substituted Query\\
    \midrule
    What variables are considered on top of the value at \textbf{1 January} when calculating the value at 31 December for \textbf{government} \textbf{grants} that are included within \textbf{trade} and other \textbf{payables}? &  What variables are considered on top of the value at \colorbox{gray!20}{\texttt{[MASK]}} when calculating the value at 31 December for \colorbox{gray!20}{\texttt{[MASK]}} \colorbox{gray!20}{\texttt{[MASK]}} that are included within \colorbox{gray!20}{\texttt{[MASK]}} and other \colorbox{gray!20}{\texttt{[MASK]}}? & What variables are considered on top of the value at \colorbox{gray!20}{New Year's Day} when calculating the value at 31 December for \colorbox{gray!20}{public} \colorbox{gray!20}{subsidies} that are included within \colorbox{gray!20}{commerce} and other \colorbox{gray!20}{liabilities}? \\ 
    \bottomrule
  \end{tabularx}}
  \caption{A comparative example of three query perturbation types.}
  \label{tab:masked query}
  
\end{table*}
\begin{table*}[ht]
  \centering
  \resizebox{0.9\linewidth}{!}{
  \begin{tabular}{lccccc}
    \toprule
Model & Perturbation Method & Hit@1&Hit@4&Hit@10&MAP@10
\\
\midrule
 \multirow{5}*{BM25}& original &28.51& 46.18 & 57.43 & 37.46 \\ \cmidrule{2-6}
 & \multirow{2}*{masked} & 0.40&6.65&11.29&3.17 \\
 & & {\small ($\downarrow$28.11)}& {\small ($\downarrow$39.53)}&{\small ($\downarrow$46.14)}&{\small ($\downarrow$34.29)}\\  \cmidrule{2-6}
 & \multirow{2}*{substituted} &5.85& 12.10& 16.13&8.94\\
 & & {\small ($\downarrow$22.66)}& {\small ($\downarrow$34.08)}&{\small ($\downarrow$41.30)}&{\small ($\downarrow$28.52)} \\
 \midrule
  \multirow{5}*{Qwen2-1.5B}& original & 23.69 & 41.37 & 53.82 & 32.84 \\ \cmidrule{2-6}
 & \multirow{2}*{masked} &2.41&4.62&5.82&3.31 \\
 & & {\small ($\downarrow$21.28)}& {\small ($\downarrow$36.75)}&{\small ($\downarrow$48.00)}&{\small ($\downarrow$29.53)}\\  \cmidrule{2-6}
 & \multirow{2}*{substituted}& 8.87& 17.14& 24.40&13.08\\
 & & {\small ($\downarrow$14.82)}& {\small ($\downarrow$24.23)}&{\small ($\downarrow$29.42)}&{\small ($\downarrow$19.76)}\\
\midrule
 \multirow{5}*{Qwen2-1.5B + BMEmbed}& original & 26.31 & 45.38 & 57.03&36.21 \\  \cmidrule{2-6}
 &\multirow{2}*{masked} & 2.21 &4.42&8.63&3.76\\
 & & {\small ($\downarrow$24.10)}& {\small ($\downarrow$40.96)}&{\small ($\downarrow$48.40)}&{\small ($\downarrow$32.45)}\\  \cmidrule{2-6}
 & \multirow{2}*{substituted} & 9.27& 18.95& 26.81&14.30\\
 & & {\small ($\downarrow$17.04)}& {\small ($\downarrow$26.43)}&{\small ($\downarrow$30.22)}&{\small ($\downarrow$21.91)}\\
 \bottomrule
  \end{tabular}}
  \caption{Controlled Retrieval Experiments with Query Perturbations on Finance-RAG.}
  \label{tab:comparative experiment}
\end{table*}

\subsection{Ablation Study of Listwise Fine-Tuning with Varying Temperatures}\label{effect_Temperatures}
We examine the effect of varying temperatures $\alpha$. For convenience, we work with its reciprocal, \(1/\alpha\), with values of 0.1, 0.2, 0.5, 0.7, and 1.0. We set $k$=500, $m$=10, and adopt the fine-to-coarse sampling strategy.

Figure \ref{tab:retrieval and temperature} shows the trend between MAP@10 and fine-tuning with different \(1/\alpha\), with the corresponding alignment and uniformity results shown in Figure \ref{tab:alpha}. Our analysis shows that \textbf{smaller temperature achieve better retrieval performance by fostering good uniformity in the embedding distribution.} In contrast, as temperature increases, uniformity decreases, even lowering it compared to the base model. This is because the higher temperature smooths the label distribution, which diminishes the distinction between learning samples and causes the embeddings to become overly clustered. Such clustering may hurt the performance of downstream tasks which require clear distinction between embeddings, as observed in our experiments, where it led to a degradation in retrieval performance.

\subsection{BMEmbed Balances Alignment and Uniformity Optimization}
Our ablation experiment and analysis have demonstrated that using the fine-to-coarse strategy with a smaller temperature is an effective way to leverage BM25, supported by both theoretical reasoning and practical results. Since main experiment we conducted in Section \ref{main experiment} is based on this strategy, here we report the uniformity and alignment of corresponding fine-tuned embedding models in Table \ref{main results 2} for further analysis.

\textbf{Embedding models fine-tuned with BMEmbed achieve better retrieval results due to increased uniformity compared to the base model, while maintaining relatively low alignment}. Comparing with CL with in-batch negatives, we observe that although uniformity has increased significantly, it does not effectively maintain or improve the alignment of the base model. This imbalance leads to instability in retrieval performance, and in some cases, even performance degradation. Specifically, we identify the ideal optimization direction, as indicated by the red arrow in the in Figure \ref{tab:sampling strategy}. BMEmbed achieves this theoretical direction on both Multihop-RAG and Finance-RAG, demonstrating its potential to balance the optimization of both uniformity and alignment.

\section{How Does BM25 Signal Boost Embedding? Integrating Lexical Sensitivity with Semantic Generalization}

While BMEmbed is fine-tuned using weak supervision signals derived from BM25, it remains unclear what specific capabilities this adaptation imparts to the embedding model. To investigate this, we design a set of controlled experiments and introduce a two-part decomposition of model behavior: \textbf{semantic generalization}, defined as the ability to capture general semantic patterns, and \textbf{lexical sensitivity}, defined as the sensitivity to domain-specific lexical cues. To assess how BMEmbed balances these two aspects, we conduct three groups of query perturbation experiments: (1) retrieval using original queries, (2) retrieval with domain-specific keywords masked, and (3) retrieval with query keywords substituted by synonyms. 

The experimental pipeline involves three stages. First, we use an LLM to extract domain-specific keywords and generate semantically appropriate synonyms for each query (see prompt details in Appendix~\ref{sec:keywords prompt}). Next, we create two perturbed versions of each query by either masking the identified keywords or substituting them with their corresponding synonyms (examples shown in Table~\ref{tab:masked query}). Finally, we evaluate model performance across these variants to assess the impact of each perturbation type. The evaluation is conducted on the Finance-RAG dataset using three methods: BM25, the base Qwen2-1.5B embedding model, and Qwen2-1.5B fine-tuned with BMEmbed. As shown in Table \ref{tab:comparative experiment}, the results provide several insights of BMEmbed:

1) \textbf{Semantic Generalization}: Compared to BM25, BMEmbed exhibits significantly less performance drop under synonym substitution (Hit@10 drop: 30.22 vs. 41.30), indicating stronger semantic generalization. Notably, even when compared to the base Qwen2-1.5B model, BMEmbed achieves slightly better absolute performance (Hit@10: 26.81 vs. 24.40) despite experiencing a similar level of performance drop (Hit@10 drop: 30.22 vs. 29.42). This suggests that our fine-tuning process not only preserves but also enhances the model’s semantic generalization ability.

2) \textbf{Lexical Sensitivity}: Under keyword masking, BMEmbed shows a larger performance drop than the base model (Hit@4 drop: 40.96 vs. 36.75), implying that BMEmbed has become more sensitive to domain-specific lexical cues, especially to the high ranking items. This indicates that while BMEmbed preserves semantic understanding, it also better incorporates keyword-level information.

These results suggest that BMEmbed effectively combines the strengths of both lexical and semantic information. This dual capability makes it particularly well-suited for domains that require adaptation to specialized terminology, or for proprietary enterprise datasets.

\section{Conclusion}
With the growing adoption of AI in real-world applications, particularly RAG systems, adapting general-purpose models to domain-specific data remains a critical challenge. In this paper, we present BMEmbed, a novel method for adapting text embedding models to private datasets (e.g., company-specific proprietary data). Since private datasets often contain specialized terminology and domain-specific language, we leverage keyword-based retrieval as a supervisory signal to fine-tune general-purpose embedding models.
Experimental results demonstrate that BMEmbed effectively enhances retrieval performance, producing more accurate query results on private datasets. As AI continues to transform industries, we hope that our proposed method can further advance the adoption and adaptation of AI in domain-specific applications, ensuring more effective and contextually relevant retrieval.

\section{Limitations}
This study has several limitations that present opportunities for future research.
First, our current method primarily focuses on the retrieval task in embedding models. However, text embeddings are also widely used in domain-specific NLP tasks such as clustering and semantic textual similarity (STS). An interesting direction for future research is exploring task-specific supervisory signals to better adapt general-purpose embedding models to private datasets for applications beyond retrieval, including clustering and STS. Second, while our method aims to develop embedding models tailored to private datasets (such as company-specific proprietary data), we evaluate it on public datasets. These datasets are chosen because they are released after the base embedding models we assess, ensuring fair comparison and public reproducibility. However, applying this method to proprietary datasets in real-world RAG scenarios remains an important next step. We hope future research will explore these practical applications to further validate and refine our approach.

\section*{Acknowledgment}
This work is partially supported by a research grant provided by HSBC. We also thank the anonymous reviewers for their thoughtful and constructive comments.

\bibliography{latex/custom}

\begin{thebibliography}{36}
\providecommand{\natexlab}[1]{#1}

\bibitem[{Anderson et~al.(2024)Anderson, Janardhanan, He, Cheng, and Flanagan}]{DBLP:conf/emnlp/AndersonJHCF24}
Peter Anderson, Mano~Vikash Janardhanan, Jason He, Wei Cheng, and Charlie Flanagan. 2024.
\newblock \href {https://aclanthology.org/2024.emnlp-industry.26} {Greenback bears and fiscal hawks: Finance is a jungle and text embeddings must adapt}.
\newblock In \emph{Proceedings of the 2024 Conference on Empirical Methods in Natural Language Processing: {EMNLP} 2024 - Industry Track, Miami, Florida, USA, November 12-16, 2024}, pages 362--370. Association for Computational Linguistics.

\bibitem[{Bachman et~al.(2019)Bachman, Hjelm, and Buchwalter}]{DBLP:conf/nips/BachmanHB19}
Philip Bachman, R.~Devon Hjelm, and William Buchwalter. 2019.
\newblock \href {https://proceedings.neurips.cc/paper/2019/hash/ddf354219aac374f1d40b7e760ee5bb7-Abstract.html} {Learning representations by maximizing mutual information across views}.
\newblock In \emph{Advances in Neural Information Processing Systems 32: Annual Conference on Neural Information Processing Systems 2019, NeurIPS 2019, December 8-14, 2019, Vancouver, BC, Canada}, pages 15509--15519.

\bibitem[{BehnamGhader et~al.(2024)BehnamGhader, Adlakha, Mosbach, Bahdanau, Chapados, and Reddy}]{DBLP:journals/corr/abs-2404-05961}
Parishad BehnamGhader, Vaibhav Adlakha, Marius Mosbach, Dzmitry Bahdanau, Nicolas Chapados, and Siva Reddy. 2024.
\newblock \href {https://doi.org/10.48550/ARXIV.2404.05961} {Llm2vec: Large language models are secretly powerful text encoders}.
\newblock \emph{arXiv preprint arXiv:2404.05961}.

\bibitem[{Cao et~al.(2007)Cao, Qin, Liu, Tsai, and Li}]{DBLP:conf/icml/CaoQLTL07}
Zhe Cao, Tao Qin, Tie{-}Yan Liu, Ming{-}Feng Tsai, and Hang Li. 2007.
\newblock \href {https://doi.org/10.1145/1273496.1273513} {Learning to rank: from pairwise approach to listwise approach}.
\newblock In \emph{Machine Learning, Proceedings of the Twenty-Fourth International Conference {(ICML} 2007), Corvallis, Oregon, USA, June 20-24, 2007}, volume 227 of \emph{{ACM} International Conference Proceeding Series}, pages 129--136. {ACM}.

\bibitem[{Chen et~al.(2024)Chen, Xiao, Zhang, Luo, Lian, and Liu}]{DBLP:conf/acl/ChenXZLLL24}
Jianlyu Chen, Shitao Xiao, Peitian Zhang, Kun Luo, Defu Lian, and Zheng Liu. 2024.
\newblock \href {https://doi.org/10.18653/V1/2024.FINDINGS-ACL.137} {M3-embedding: Multi-linguality, multi-functionality, multi-granularity text embeddings through self-knowledge distillation}.
\newblock In \emph{Findings of the Association for Computational Linguistics, {ACL} 2024, Bangkok, Thailand and virtual meeting, August 11-16, 2024}, pages 2318--2335. Association for Computational Linguistics.

\bibitem[{Chen et~al.(2020)Chen, Kornblith, Norouzi, and Hinton}]{DBLP:conf/icml/ChenK0H20}
Ting Chen, Simon Kornblith, Mohammad Norouzi, and Geoffrey~E. Hinton. 2020.
\newblock \href {http://proceedings.mlr.press/v119/chen20j.html} {A simple framework for contrastive learning of visual representations}.
\newblock In \emph{Proceedings of the 37th International Conference on Machine Learning, {ICML} 2020, 13-18 July 2020, Virtual Event}, volume 119 of \emph{Proceedings of Machine Learning Research}, pages 1597--1607. {PMLR}.

\bibitem[{Cormack et~al.(2009)Cormack, Clarke, and B{\"{u}}ttcher}]{DBLP:conf/sigir/CormackCB09}
Gordon~V. Cormack, Charles L.~A. Clarke, and Stefan B{\"{u}}ttcher. 2009.
\newblock \href {https://doi.org/10.1145/1571941.1572114} {Reciprocal rank fusion outperforms condorcet and individual rank learning methods}.
\newblock In \emph{Proceedings of the 32nd Annual International {ACM} {SIGIR} Conference on Research and Development in Information Retrieval, {SIGIR} 2009, Boston, MA, USA, July 19-23, 2009}, pages 758--759. {ACM}.

\bibitem[{Dehghani et~al.(2017)Dehghani, Zamani, Severyn, Kamps, and Croft}]{DBLP:conf/sigir/DehghaniZSKC17}
Mostafa Dehghani, Hamed Zamani, Aliaksei Severyn, Jaap Kamps, and W.~Bruce Croft. 2017.
\newblock \href {https://doi.org/10.1145/3077136.3080832} {Neural ranking models with weak supervision}.
\newblock In \emph{Proceedings of the 40th International {ACM} {SIGIR} Conference on Research and Development in Information Retrieval, Shinjuku, Tokyo, Japan, August 7-11, 2017}, pages 65--74. {ACM}.

\bibitem[{Devlin et~al.(2019)Devlin, Chang, Lee, and Toutanova}]{DBLP:conf/naacl/DevlinCLT19}
Jacob Devlin, Ming{-}Wei Chang, Kenton Lee, and Kristina Toutanova. 2019.
\newblock \href {https://doi.org/10.18653/V1/N19-1423} {{BERT:} pre-training of deep bidirectional transformers for language understanding}.
\newblock In \emph{Proceedings of the 2019 Conference of the North American Chapter of the Association for Computational Linguistics: Human Language Technologies, {NAACL-HLT} 2019, Minneapolis, MN, USA, June 2-7, 2019, Volume 1 (Long and Short Papers)}, pages 4171--4186. Association for Computational Linguistics.

\bibitem[{Gao et~al.(2021)Gao, Yao, and Chen}]{DBLP:conf/emnlp/GaoYC21}
Tianyu Gao, Xingcheng Yao, and Danqi Chen. 2021.
\newblock \href {https://doi.org/10.18653/V1/2021.EMNLP-MAIN.552} {Simcse: Simple contrastive learning of sentence embeddings}.
\newblock In \emph{Proceedings of the 2021 Conference on Empirical Methods in Natural Language Processing, {EMNLP} 2021, Virtual Event / Punta Cana, Dominican Republic, 7-11 November, 2021}, pages 6894--6910. Association for Computational Linguistics.

\bibitem[{Gao et~al.(2023)Gao, Xiong, Gao, Jia, Pan, Bi, Dai, Sun, Guo, Wang, and Wang}]{DBLP:journals/corr/abs-2312-10997}
Yunfan Gao, Yun Xiong, Xinyu Gao, Kangxiang Jia, Jinliu Pan, Yuxi Bi, Yi~Dai, Jiawei Sun, Qianyu Guo, Meng Wang, and Haofen Wang. 2023.
\newblock \href {https://doi.org/10.48550/ARXIV.2312.10997} {Retrieval-augmented generation for large language models: {A} survey}.
\newblock \emph{arXiv preprint arXiv:2312.10997}.

\bibitem[{Haddad and Ghosh(2019)}]{DBLP:conf/sigir/HaddadG19}
Dany Haddad and Joydeep Ghosh. 2019.
\newblock \href {https://doi.org/10.1145/3331184.3331272} {Learning more from less: Towards strengthening weak supervision for ad-hoc retrieval}.
\newblock In \emph{Proceedings of the 42nd International {ACM} {SIGIR} Conference on Research and Development in Information Retrieval, {SIGIR} 2019, Paris, France, July 21-25, 2019}, pages 857--860. {ACM}.

\bibitem[{Hjelm et~al.(2019)Hjelm, Fedorov, Lavoie{-}Marchildon, Grewal, Bachman, Trischler, and Bengio}]{DBLP:conf/iclr/HjelmFLGBTB19}
R.~Devon Hjelm, Alex Fedorov, Samuel Lavoie{-}Marchildon, Karan Grewal, Philip Bachman, Adam Trischler, and Yoshua Bengio. 2019.
\newblock \href {https://openreview.net/forum?id=Bklr3j0cKX} {Learning deep representations by mutual information estimation and maximization}.
\newblock In \emph{7th International Conference on Learning Representations, {ICLR} 2019, New Orleans, LA, USA, May 6-9, 2019}. OpenReview.net.

\bibitem[{Hu et~al.(2022)Hu, Shen, Wallis, Allen{-}Zhu, Li, Wang, Wang, and Chen}]{DBLP:conf/iclr/HuSWALWWC22}
Edward~J. Hu, Yelong Shen, Phillip Wallis, Zeyuan Allen{-}Zhu, Yuanzhi Li, Shean Wang, Lu~Wang, and Weizhu Chen. 2022.
\newblock \href {https://openreview.net/forum?id=nZeVKeeFYf9} {Lora: Low-rank adaptation of large language models}.
\newblock In \emph{The Tenth International Conference on Learning Representations, {ICLR} 2022, Virtual Event, April 25-29, 2022}. OpenReview.net.

\bibitem[{Izacard et~al.(2022)Izacard, Caron, Hosseini, Riedel, Bojanowski, Joulin, and Grave}]{DBLP:journals/tmlr/IzacardCHRBJG22}
Gautier Izacard, Mathilde Caron, Lucas Hosseini, Sebastian Riedel, Piotr Bojanowski, Armand Joulin, and Edouard Grave. 2022.
\newblock \href {https://openreview.net/forum?id=jKN1pXi7b0} {Unsupervised dense information retrieval with contrastive learning}.
\newblock \emph{Trans. Mach. Learn. Res.}, 2022.

\bibitem[{Karpukhin et~al.(2020)Karpukhin, Oguz, Min, Lewis, Wu, Edunov, Chen, and Yih}]{DBLP:conf/emnlp/KarpukhinOMLWEC20}
Vladimir Karpukhin, Barlas Oguz, Sewon Min, Patrick S.~H. Lewis, Ledell Wu, Sergey Edunov, Danqi Chen, and Wen{-}tau Yih. 2020.
\newblock \href {https://doi.org/10.18653/V1/2020.EMNLP-MAIN.550} {Dense passage retrieval for open-domain question answering}.
\newblock In \emph{Proceedings of the 2020 Conference on Empirical Methods in Natural Language Processing, {EMNLP} 2020, Online, November 16-20, 2020}, pages 6769--6781. Association for Computational Linguistics.

\bibitem[{Lee et~al.(2024)Lee, Roy, Xu, Raiman, Shoeybi, Catanzaro, and Ping}]{DBLP:journals/corr/abs-2405-17428}
Chankyu Lee, Rajarshi Roy, Mengyao Xu, Jonathan Raiman, Mohammad Shoeybi, Bryan Catanzaro, and Wei Ping. 2024.
\newblock \href {https://doi.org/10.48550/ARXIV.2405.17428} {Nv-embed: Improved techniques for training llms as generalist embedding models}.
\newblock \emph{arXiv preprint arXiv:2405.17428}.

\bibitem[{Li et~al.(2023)Li, Zhang, Zhang, Long, Xie, and Zhang}]{DBLP:journals/corr/abs-2308-03281}
Zehan Li, Xin Zhang, Yanzhao Zhang, Dingkun Long, Pengjun Xie, and Meishan Zhang. 2023.
\newblock \href {https://doi.org/10.48550/ARXIV.2308.03281} {Towards general text embeddings with multi-stage contrastive learning}.
\newblock \emph{arXiv preprint arXiv:2308.03281}.

\bibitem[{Liu et~al.(2024)Liu, Yang, and Tam}]{liu2024beyond}
Jiaxin Liu, Yi~Yang, and Kar~Yan Tam. 2024.
\newblock Beyond surface similarity: Detecting subtle semantic shifts in financial narratives.
\newblock In \emph{Findings of the Association for Computational Linguistics: NAACL 2024}, pages 2641--2652.

\bibitem[{Moreira et~al.(2024)Moreira, Osmulski, Xu, Ak, Schifferer, and Oldridge}]{moreira2024nv}
Gabriel de Souza~P Moreira, Radek Osmulski, Mengyao Xu, Ronay Ak, Benedikt Schifferer, and Even Oldridge. 2024.
\newblock Nv-retriever: Improving text embedding models with effective hard-negative mining.
\newblock \emph{arXiv preprint arXiv:2407.15831}.

\bibitem[{Muennighoff et~al.(2023)Muennighoff, Tazi, Magne, and Reimers}]{DBLP:conf/eacl/MuennighoffTMR23}
Niklas Muennighoff, Nouamane Tazi, Lo{\"{\i}}c Magne, and Nils Reimers. 2023.
\newblock \href {https://doi.org/10.18653/V1/2023.EACL-MAIN.148} {{MTEB:} massive text embedding benchmark}.
\newblock In \emph{Proceedings of the 17th Conference of the European Chapter of the Association for Computational Linguistics, {EACL} 2023, Dubrovnik, Croatia, May 2-6, 2023}, pages 2006--2029. Association for Computational Linguistics.

\bibitem[{Ni et~al.(2022)Ni, {\'{A}}brego, Constant, Ma, Hall, Cer, and Yang}]{DBLP:conf/acl/NiACMHCY22}
Jianmo Ni, Gustavo~Hern{\'{a}}ndez {\'{A}}brego, Noah Constant, Ji~Ma, Keith~B. Hall, Daniel Cer, and Yinfei Yang. 2022.
\newblock \href {https://doi.org/10.18653/V1/2022.FINDINGS-ACL.146} {Sentence-t5: Scalable sentence encoders from pre-trained text-to-text models}.
\newblock In \emph{Findings of the Association for Computational Linguistics: {ACL} 2022, Dublin, Ireland, May 22-27, 2022}, pages 1864--1874. Association for Computational Linguistics.

\bibitem[{Pipitone and Alami(2024)}]{DBLP:journals/corr/abs-2408-10343}
Nicholas Pipitone and Ghita~Houir Alami. 2024.
\newblock \href {https://doi.org/10.48550/ARXIV.2408.10343} {Legalbench-rag: {A} benchmark for retrieval-augmented generation in the legal domain}.
\newblock \emph{arXiv preprint arXiv:2408.10343}.

\bibitem[{Raffel et~al.(2020)Raffel, Shazeer, Roberts, Lee, Narang, Matena, Zhou, Li, and Liu}]{DBLP:journals/jmlr/RaffelSRLNMZLL20}
Colin Raffel, Noam Shazeer, Adam Roberts, Katherine Lee, Sharan Narang, Michael Matena, Yanqi Zhou, Wei Li, and Peter~J. Liu. 2020.
\newblock \href {https://jmlr.org/papers/v21/20-074.html} {Exploring the limits of transfer learning with a unified text-to-text transformer}.
\newblock \emph{J. Mach. Learn. Res.}, 21:140:1--140:67.

\bibitem[{Reimers and Gurevych(2019)}]{DBLP:conf/emnlp/ReimersG19}
Nils Reimers and Iryna Gurevych. 2019.
\newblock \href {https://doi.org/10.18653/V1/D19-1410} {Sentence-bert: Sentence embeddings using siamese bert-networks}.
\newblock In \emph{Proceedings of the 2019 Conference on Empirical Methods in Natural Language Processing and the 9th International Joint Conference on Natural Language Processing, {EMNLP-IJCNLP} 2019, Hong Kong, China, November 3-7, 2019}, pages 3980--3990. Association for Computational Linguistics.

\bibitem[{Robertson and Zaragoza(2009)}]{DBLP:journals/ftir/RobertsonZ09}
Stephen~E. Robertson and Hugo Zaragoza. 2009.
\newblock \href {https://doi.org/10.1561/1500000019} {The probabilistic relevance framework: {BM25} and beyond}.
\newblock \emph{Found. Trends Inf. Retr.}, 3(4):333--389.

\bibitem[{Tan et~al.(2022)Tan, Shao, Wu, Yang, and Song}]{DBLP:conf/acl/TanSWYS22}
Haochen Tan, Wei Shao, Han Wu, Ke~Yang, and Linqi Song. 2022.
\newblock \href {https://doi.org/10.18653/V1/2022.FINDINGS-ACL.22} {A sentence is worth 128 pseudo tokens: {A} semantic-aware contrastive learning framework for sentence embeddings}.
\newblock In \emph{Findings of the Association for Computational Linguistics: {ACL} 2022, Dublin, Ireland, May 22-27, 2022}, pages 246--256. Association for Computational Linguistics.

\bibitem[{Tang and Yang(2024{\natexlab{a}})}]{DBLP:journals/corr/abs-2409-18511}
Yixuan Tang and Yi~Yang. 2024{\natexlab{a}}.
\newblock \href {https://doi.org/10.48550/ARXIV.2409.18511} {Do we need domain-specific embedding models? an empirical investigation}.
\newblock \emph{arXiv preprint arXiv:2409.18511}.

\bibitem[{Tang and Yang(2024{\natexlab{b}})}]{DBLP:journals/corr/abs-2401-15391}
Yixuan Tang and Yi~Yang. 2024{\natexlab{b}}.
\newblock \href {https://doi.org/10.48550/ARXIV.2401.15391} {Multihop-rag: Benchmarking retrieval-augmented generation for multi-hop queries}.
\newblock \emph{arXiv preprint arXiv:2401.15391}.

\bibitem[{Tang and Yang(2024{\natexlab{c}})}]{tang2024pooling}
Yixuan Tang and Yi~Yang. 2024{\natexlab{c}}.
\newblock \href {https://arxiv.org/abs/2409.02727} {Pooling and attention: What are effective designs for llm-based embedding models?}
\newblock \emph{Preprint}, arXiv:2409.02727.

\bibitem[{van~den Oord et~al.(2018)van~den Oord, Li, and Vinyals}]{DBLP:journals/corr/abs-1807-03748}
A{\"{a}}ron van~den Oord, Yazhe Li, and Oriol Vinyals. 2018.
\newblock \href {https://arxiv.org/abs/1807.03748} {Representation learning with contrastive predictive coding}.
\newblock \emph{arXiv preprint arXiv:1807.03748}.

\bibitem[{Wang et~al.(2022)Wang, Yang, Huang, Jiao, Yang, Jiang, Majumder, and Wei}]{DBLP:journals/corr/abs-2212-03533}
Liang Wang, Nan Yang, Xiaolong Huang, Binxing Jiao, Linjun Yang, Daxin Jiang, Rangan Majumder, and Furu Wei. 2022.
\newblock \href {https://doi.org/10.48550/ARXIV.2212.03533} {Text embeddings by weakly-supervised contrastive pre-training}.
\newblock \emph{arXiv preprint arXiv:2212.03533}.

\bibitem[{Wang and Isola(2020)}]{DBLP:conf/icml/0001I20}
Tongzhou Wang and Phillip Isola. 2020.
\newblock \href {http://proceedings.mlr.press/v119/wang20k.html} {Understanding contrastive representation learning through alignment and uniformity on the hypersphere}.
\newblock In \emph{Proceedings of the 37th International Conference on Machine Learning, {ICML} 2020, 13-18 July 2020, Virtual Event}, volume 119 of \emph{Proceedings of Machine Learning Research}, pages 9929--9939. {PMLR}.

\bibitem[{Xia et~al.(2008)Xia, Liu, Wang, Zhang, and Li}]{xia2008listwise}
Fen Xia, Tie-Yan Liu, Jue Wang, Wensheng Zhang, and Hang Li. 2008.
\newblock Listwise approach to learning to rank: theory and algorithm.
\newblock In \emph{Proceedings of the 25th international conference on Machine learning}, pages 1192--1199.

\bibitem[{Xin et~al.(2022)Xin, Xiong, Srinivasan, Sharma, Jose, and Bennett}]{DBLP:conf/acl/XinX0SJ022}
Ji~Xin, Chenyan Xiong, Ashwin Srinivasan, Ankita Sharma, Damien Jose, and Paul Bennett. 2022.
\newblock \href {https://doi.org/10.18653/V1/2022.FINDINGS-ACL.316} {Zero-shot dense retrieval with momentum adversarial domain invariant representations}.
\newblock In \emph{Findings of the Association for Computational Linguistics: {ACL} 2022, Dublin, Ireland, May 22-27, 2022}, pages 4008--4020. Association for Computational Linguistics.

\bibitem[{Zhou et~al.(2022)Zhou, Zhang, Zhao, and Wen}]{DBLP:conf/acl/ZhouZZW22}
Kun Zhou, Beichen Zhang, Wayne~Xin Zhao, and Ji{-}Rong Wen. 2022.
\newblock \href {https://doi.org/10.18653/V1/2022.ACL-LONG.423} {Debiased contrastive learning of unsupervised sentence representations}.
\newblock In \emph{Proceedings of the 60th Annual Meeting of the Association for Computational Linguistics (Volume 1: Long Papers), {ACL} 2022, Dublin, Ireland, May 22-27, 2022}, pages 6120--6130. Association for Computational Linguistics.

\end{thebibliography}

\appendix

\section{Prompts Used for Domain Query Generation} \label{sec:appendix_prompt}
The LLM prompts used in the domain query generation stage are detailed as follows:

{\small
\begin{center}
\fcolorbox{black}{gray!10}{\parbox{.96\linewidth}{
\textbf{Event Extraction Prompt:}

Given a document, please extract all the events and their associated topics and organization in the context.

Note: 
1. The event should not contain ambiguous references, such as ’he’,’ she,’ and ’it’, and should use complete names. \\
2. You should give at least one passage in the original text associated to the event you extract, DO NOT make up any event.\\
3. If there are multiple paragraphs associated to the extracted event, please list and number all of them.\\
4. If the event does not contain some of the arguments mentioned above, please leave it empty.\\
5. The type of Event involves fine-grained events and general events, where fine-grained events focus on specific facts and details while general events are summarizations of happened fine-grained events.\\
6. Please return the fine-grained events first, then return general events.\\
The document is:\\
\{doc\}\\
Please return the extracted event in the following format with following arguments:

[Event]:

[Topic]:

[Original context]: 1. ......\\
2. ......\\
...

[Type]:

Events you extract are:

}}
\end{center}
}
{\small
\begin{center}
\fcolorbox{black}{gray!10}{\parbox{.96\linewidth}{
\textbf{Query Synthesis Prompt:}

Given several events and their original source document, please ask several questions according to the infomation and give the original reference paragraph following this format:

[Envent]:

[Question]:

Note: 1. Don't need to mention all the arguments in your question. 

2. You can involve the original document information, but make sure that your question is about the topic of the given event.

3. You should ask questions separately to different events.

Document:

\{doc\}

Event:

\{event\}

Your question towards given event:

}}
\end{center}
}

\section{Case Study of Query Generation}\label{case_query_generation}
In this section, we present a real query generation process, showcasing the input document, intermediate extracted events, and the final generated query.
{\small
\begin{center}
\fcolorbox{black}{gray!10}{\parbox{.96\linewidth}{
\textbf{Document Chunk from Multihop-RAG:}\\
Table of Contents Table of Contents Echo, Fire TV, and Kindle deals Apple deals TV deals Laptop deals Headphone and earbud deals Tablet deals Gaming deals Speaker deals Vacuum deals Kitchen deals Smart home deals Fitness deals Beauty tech deals Drone deals Camera deals Lego deals Gift card deals UPDATE: Nov. 27, 2023, 5:00 a.m. EST This post has been updated with all of the latest Cyber Monday deals available at Amazon. Amazon is dragging out the year's biggest shopping holiday(s) into 11 days of deals. The retail giant began its Black Friday sale in the early morning of Friday, Nov. 17 (a week ahead of schedule) and was on top of making the switch to Cyber Monday language in the wee hours of Saturday, Nov. 25. Official Cyber Monday mode, which is currently on through Monday, Nov. 27, includes both a ton of deals carried over from Black Friday plus some new ones.}}\\
\end{center}
}

{\small
\begin{center}
\fcolorbox{black}{gray!10}{\parbox{.96\linewidth}{
\textbf{GPT-4o Extracted Events:}\\\#\#\# Fine-Grained Events 

1. 
[Event]: Amazon began its Black Friday sale. 

[Topic]: Black Friday sale

[Original context]: "The retail giant began its Black Friday sale in the early morning of Friday, Nov. 17 (a week ahead of schedule)..." 

[Type]: Fine-grained

2. 
[Event]: Amazon switched to Cyber Monday language. 

[Topic]: Cyber Monday sale 

[Original context]: "...and was on top of making the switch to Cyber Monday language in the wee hours of Saturday, Nov. 25." 

[Type]: Fine-grained

3. 
[Event]: Amazon's Cyber Monday mode includes deals carried over from Black Friday plus some new ones. 

[Topic]: Cyber Monday sale 

[Original context]: "Official Cyber Monday mode, which is currently on through Monday, Nov. 27, includes both a ton of deals carried over from Black Friday plus some new ones." 

[Type]: Fine-grained }}
\end{center}
}

{\small
\begin{center}
\fcolorbox{black}{gray!10}{\parbox{.96\linewidth}{
\textbf{GPT-4o-mini Generated Query:} \\
1. [Event]: Amazon began its Black Friday sale. 
[Question]: How did Amazon's early start to the Black Friday sale impact customer engagement compared to previous years? 

2. [Event]: Amazon switched to Cyber Monday language. 
[Question]: What strategies did Amazon employ to transition from Black Friday to Cyber Monday promotions? 

3. [Event]: Amazon's Cyber Monday mode includes deals carried over from Black Friday plus some new ones. [Question]: What types of new deals can customers expect during Amazon's Cyber Monday sale compared to those from Black Friday? 
}}
\end{center}
}

\section{Experimental Details for Generalization Settings}
\label{appendix:generalization}
\paragraph{Setting 1:} We use the all-MiniLM-L6-v2 model, which has only 22.7M parameters. We fine-tune it on the Multihop-RAG dataset using a learning rate of 2e-5 for 500 steps, with the same sampling strategy used in our main experiments for this dataset. The full fine-tuning process required only 24 minutes on a single NVIDIA 3090 GPU, with a peak memory usage of 4GB. The results are presented in the Table \ref{tab:minilm}.
\paragraph{Setting 2:} We conducted an additional experiment using ListMLE \cite{xia2008listwise} on the Qwen2-1.5B model and the Multihop-RAG dataset, under the same settings as our main experiment. This setup is compared against ListNet-based training used in the our main experiment. The results are presented in Table~\ref{tab:listmle}.
\paragraph{Setting 3:} We conducted an additional experiment to evaluate the adapted embeddings on a semantic similarity task. Specifically, we used the model fine-tuned on the Finance-RAG dataset and evaluated it on the FinSTS dataset \cite{liu2024beyond}, a well-annotated benchmark designed to detect subtle semantic shifts in financial narratives. Since both datasets are based on financial reports, FinSTS serves as a "private evaluation set" in this context. In this evaluation, we adopted a last-token pooling configuration and used the Cosine Spearman Correlation as the evaluation metric. The results are presented in Table~\ref{tab:sts}.

\begin{table}[t]
  \centering
  \resizebox{1\linewidth}{!}{
  \begin{tabular}{lcccc}
    \toprule
    Model & Hit@1 & Hit@4 & Hit@10 & MAP@10 \\
    \midrule
    Qwen2-1.5B & 33.97 & 59.69 & 76.50 & 22.22 \\
    e5-mistral-7B & 29.49 & 54.99 & 75.39 & 20.33 \\
    MiniLM  & 17.52 & 39.96 & 55.79 & 12.55 \\
    \multirow{2}*{MiniLM+BMEmbed} & 32.77 & 60.18 & 78.27 & 22.40 \\
    & (\small{+15.25})&(\small{+20.22})&(\small{+22.48})&(\small{+9.85}) \\
    \bottomrule
  \end{tabular}}
  \caption{Retrieval Performance of Different Models on MultihopRAG.}
  \label{tab:minilm}
\end{table}
\begin{table}[t]
  \centering
  \resizebox{0.8\linewidth}{!}{
  \begin{tabular}{lcccc}
    \toprule
    Method & Hit@1 & Hit@4 & Hit@10 & MAP@10 \\
    \midrule
    ListNet & 40.58 & 68.34 & 	40.58 & 26.54 \\
    ListMLE  & 39.87 & 67.98 & 83.10 & 26.29 \\

    \bottomrule
  \end{tabular}}
  \caption{ListMLE vs. ListNet under identical training settings.}
  \label{tab:listmle}
\end{table}
\begin{table}[t]
  \centering
  \resizebox{1.0\linewidth}{!}{
  \begin{tabular}{lccc}
    \toprule
    Model & without BMEmbed	 & with BMEmbed	 &Improvement \\
    \midrule
    Qwen2-1.5B & 0.2566 & 0.2727 & 	+0.0161 \\
    e5-mistral-7B  & 0.2678 & 0.3024 & +0.0346 \\
    \bottomrule
  \end{tabular}}
  \caption{Evaluation on FinSTS with Cosine Spearman correlation.}
  \label{tab:sts}
\end{table}

\section{Ablation Study}\label{app:ablation}
\subsection{Ablation Study of Query Generation Module}
We conduct experiments to investigate the impact of the number of synthetic queries used for fine-tuning. Specifically, we compare three settings: (1) using the full set of synthetic queries, (2) using a randomly sampled 50\% subset, and (3) using a randomly sampled 25\% subset. To control for the total number of training samples, we change the number of listwise samples generated per query. Specifically, we increase the number of sampled ranking lists per query accordingly when using fewer queries, ensuring the overall amount of training data remains constant. All experiments are conducted on the Multihop-RAG dataset using the Qwen2-1.5B model. All other settings are kept fixed, including the sampling strategy, number of training steps (1,000), and the temperature (1.0) used in listwise fine-tuning. 

As shown in Tabel \ref{tab:query num}, no significant performance difference is observed across the three settings, suggesting that the number of synthetic queries has limited impact on the model’s performance. This indicates that BMEmbed can compensate for fewer queries by generating multiple listwise samples per query, thereby maintaining training signal quality.

\begin{table*}[t!]
  \centering
  \resizebox{0.8\linewidth}{!}{ \begin{tabular}{lccccccc}
    \toprule
    Setting &Samples per Query& Total Samples&Hit@1 & Hit@4 & Hit@10 & MAP@10 \\
    \midrule
    full set &1 & 5,972 & 41.02 & 69.36 & 84.79 &26.96 \\

    subset(50\%) & 2 & 5,972 & 39.91 & 68.43 & 84.21 &26.30 \\

    subset(25\%) & 4 & 5,972 & 40.31 & 68.03 & 84.08 &26.48 \\
    
      \bottomrule
      
  \end{tabular}
}
  \caption{Ablation study of Query Generation Module.}
  \label{tab:query num}
\end{table*}

\subsection{Ablation Study of Relevant Sampling Module}
We conduct three sets of experiments on Multihop-RAG and Qwen model while controlling different variables, investigating four key factors according to our pipeline:
\begin{itemize}
    \item selection of $k$, we explore values of  \(k\) at 200, 500, and 1000;
    \item selection of $m$, we examine \(m\) values ranging from 6 to 10;
    \item sampling strategy, compared fine-to-coarse and uniform approaches, fixing the first partition from 0 to 3 for an informative positive sample, while dividing the remaining partitions based on the chosen strategy. Specifically, when using the fine-to-coarse strategy, for a given $k$ and $m$, the length of the next interval is twice the length of the previous interval. This can be represented by the formula: \(L(\mathcal{P}_{i+1})= 2L(\mathcal{P}_{i})\);
    \item hyperparameter \(\alpha\), for convenience, we work with its reciprocal, \(1/\alpha\), with values of 0.1, 0.2, 0.5, 0.7, and 1.0. 
\end{itemize}

Our experiments are structured as follows:
\begin{enumerate}
    \item We fix temperature = 1 and $k$=1000, and conduct experiments with different values of $m$ and sampling strategies.
    \item We fix temperature = 1, $m$=10, and the fine-to-coarse strategy, then investigate different values of $k$.
    \item We fix $k$=500, $m$=10,  and the fine-to-coarse strategy, then examine the effect of varying temperature.
\end{enumerate}

\begin{table*}[t!]
  \centering
  \resizebox{\textwidth}{!}{ \begin{tabular}{lcccccc}
    \toprule
    Method & Alignment & Uniformity & Hit@1 & Hit@4 & Hit@10 & MAP@10 \\
    \midrule
 Base &  1.2422 &2.7624 & 33.97&59.69&76.50&22.22\\
 m=6 k=1000 fine-to-coarse & 1.2031& 3.1258& 38.94 & 67.45& 82.44&25.94\\
 m=7 k=1000 fine-to-coarse & 1.1953& 3.1907& 41.02 &69.00& 83.99&26.76\\
 m=8 k=1000 fine-to-coarse & 1.1953&3.3276& 39.38&68.91& 83.33&26.29\\
 m=9 k=1000 fine-to-coarse & 1.2031& 3.3266&40.58&68.34& 83.06&26.54\\
 m=10 k=1000 fine-to-coarse &   1.2031& 3.3267& 40.04&68.43&83.55&26.43\\
 m=6 k=1000 uniform&   1.2734& 3.6012& 36.98& 64.79& 80.27&24.41\\
 m=7 k=1000 uniform&   1.2656& 3.5860&36.76& 65.19& 81.37&24.79\\
 m=8 k=1000 uniform&  1.2578& 3.6276& 38.67& 67.49&82.35&25.61\\
 m=9 k=1000 uniform&  1.2578& 3.6222& 38.18& 65.90& 81.46&25.24\\
  m=10 k=1000 uniform&  1.2734& 3.6265& 36.50&64.26&80.71&24.39\\
  \midrule
 k=1000 unifom m=10&   1.2734& 3.6265& 36.50&64.26&80.71&24.39\\
 k=500 unifom m=10&   1.2578& 3.6303& 36.76& 65.45& 81.46&24.72\\
 k=200 unifom m=10&   1.2422& 3.6452& 37.69& 66.39& 82.97&25.23\\
 k=1000 fine-to-coarse m=10&    1.2031& 3.3267& 40.04&68.43&83.55&26.43\\
 k=500 fine-to-coarse m=10&  1.1953& 3.3675& 40.71& 68.74&83.50&26.67\\
 k=200 fine-to-coarse m=10&  1.1953& 3.3896& 38.85& 68.65& 83.10&26.11\\
 \midrule
\(1/\alpha\)=0.1 k=1000 fine-to-coarse m=10&  1.1953& 2.1774& 35.48& 63.02& 78.14&23.96\\
\(1/\alpha\)=0.2 k=1000 fine-to-coarse m=10& 1.1875& 2.6560& 37.83& 66.43&81.46&25.47\\
\(1/\alpha\)=0.5 k=1000 fine-to-coarse m=10& 1.1875& 3.2849& 40.09& 67.63& 82.88&26.34\\
\(1/\alpha\)=0.7 k=1000 fine-to-coarse m=10& 1.1953&3.3411& 39.96& 68.29& 83.10&26.45\\
 \(1/\alpha\)=1.0 k=1000 fine-to-coarse m=10& 1.1953& 3.3675& 40.71& 68.74&83.50&26.67\\
  \bottomrule
  \end{tabular}
}
  \caption{Ablation study of Relevant Sampling
Module.}
  \label{tab:ablation}
\end{table*}

Our ablation experiment results in Table \ref{tab:ablation} demonstrate that, \textbf{fine-tuned embedding model with lower alignment and higher uniformity tend to achieve better result on retrieval task.} We observe a strong correlation between retrieval performance and these two properties. Specifically, embedding models with better alignment tend to achieve superior retrieval results. Moreover, when alignment is similar, models with larger uniformity exhibit better retrieval performance. This suggests that we can leverage our strategy to adjust alignment and uniformity, ultimately optimizing retrieval performance.

\section{Alignment and Uniformity: Details and Discussion}\label{app:alignment_uniformity}

In the work of \citet{DBLP:conf/icml/0001I20}, \textbf{Alignment}, which measures how well similar data points are positioned in the embedding space, is quantified by the mean Euclidean distance between the embeddings of all positive pairs. \textbf{Uniformity}, which reflects how well the data points are distributed across the embedding space, is quantified using the Gaussian potential kernel, capturing the pairwise similarity across all data points in the distribution, they are denoted as follows:
\begin{equation*}
\small
  \begin{split}
  &\text{Alignment} = \mathbb{E}_{x,y\in pos} [\left\| e(x) - e(y)\right\|_2^2] \\
&\text{Uniformity}  = log \space \mathbb{E}_{x,y\in p_{data}} [exp(-2\left\| h(x) - h(y)\right\|_2^2)]
 \end{split}
\end{equation*}
where \(x,y \in pos\) represents the positive pairs in the dataset, and \( p_{data}\) is the data distribution of all data points, \(e(\cdot)\) is the embedding model that maps input data points to their corresponding embeddings in a high-dimensional space.  In our experiments,  \(x,y \in pos\) refer to the question and its corresponding evidence chunk, while we randomly sample chunks from each document, forming a set of  \(p_{data}\) to compute uniformity.

Since fine-tuning can further amend the model's alignment \cite{DBLP:conf/emnlp/GaoYC21}, making it difficult to compare across different models, we introduce a scaling factor to address this. A model with high alignment does not necessarily perform worse in retrieval than one with low alignment. If a high-alignment model also ensures that negative samples are more dispersed relative to positive ones, it can still achieve strong retrieval performance. Considering this, we define the distance between the query and its nearest embedding in the database as a scaling factor for alignment. In the following experiments, we use the normalized version of alignment, which denotes as follows:

{\small\begin{equation*}
\text{Alignment}_{norm} = \mathbb{E}_{x,y\in pos} [\frac{\left\| e(x) - e(y)\right\|_2^2}{\left\| e(x) - e(y_{\text{nearest}})\right\|_2^2}] 
\end{equation*}
}, where \(e(y_\text{nearest})\) refers to the closest embedding in the database to the question embedding \(e(x)\).
Finally, the original uniformity is a negative value, in our experiments, we report the absolute value of uniformity. This makes comparison and analysis easier, and a larger absolute value indicates that the embedding model distribution is more uniform.

\section{Prompts Used for Query Perturbation} \label{sec:keywords prompt}
The LLM prompts used in the keywords masking experiments are detailed as follows:

{\small
\begin{center}
\fcolorbox{black}{gray!10}{\parbox{.96\linewidth}{
\textbf{Prompt for Extracting Keywords:}

Given a query and a paragraph including the answer of the query, please extract all the common keywords that query and paragraph both have:

Note:

1. The definition of keywords is: words in the query and paragraph that are particularly distinctive and related to the main topic. Less important pronouns or frequently occurring words do not fall into this category.

2. The words you extract must appear in both the query and the paragraph.

3. Do not output other format, just list all the words as follows:

investigation, Eastwood, Filing

Query:

\{query\}

Paragraph:

\{paragraph\}

keywords:

}}
\end{center}
}
{\small
\begin{center}
\fcolorbox{black}{gray!10}{\parbox{.96\linewidth}{
\textbf{Prompt for Generating Synonyms:}

Given a query and a set of its keywords, generate substituted words or phrases for these keywords that preserve the original semantic meaning of the query. 

Note:

1. Ensure the number of keywords remains unchanged, with one substitution for each keyword. Maintain the query's intent, context, and grammatical correctness. 

2. Avoid altering the overall structure and purpose of the query.

3. Return the substituted keywords in the same format with Keywords like:
investigation, Eastwood, Filing

Query:

\{query\}

Keywords:

\{keywords\}

Your substituted keywords:

}}
\end{center}
}

\end{document}